\documentclass[aps,prl,twocolumn,floats,amsmath,amssymb,superscriptaddress]{revtex4}
\usepackage{graphicx}
\usepackage{epsfig}
\usepackage{amsfonts}
\usepackage{amsmath}
\usepackage{natbib}
\usepackage{multirow}
\usepackage{bm}
\usepackage{epstopdf}
\usepackage{color}
\usepackage{ulem}
\usepackage{notes2bib}

\newcommand{\exciting}{{\usefont{T1}{lmtt}{b}{n}exciting}}
\newcommand{\LO}{{\usefont{T1}{lmtt}{b}{n}LayerOptics}}

\begin{document}
\title{\exciting\ core-level spectroscopy}
\author{Claudia Draxl}
\author{Caterina Cocchi}
\affiliation{Institut f\"{u}r Physik and IRIS Adlershof, Humboldt-Universit\"{a}t zu Berlin, Berlin, Germany}
\affiliation{European Theoretical Spectroscopic Facility (ETSF)}
\begin{abstract}
We delineate an {\it ab initio} approach for obtaining x-ray absorption spectra as provided by many-body perturbation theory, together with its realization within an all-electron framework. Employing the Bethe-Salpeter equation, 
we address excitations from different absorption edges in a broad range of material classes, analyzing exciton binding strengths and character. We also discuss when the supercell core-hole approach is likely to fail.
\end{abstract}
\maketitle

\section{Introduction} 

While density-functional theory (DFT) is the workhorse of computational materials science that concerns groundstate properties of materials, theoretical spectroscopy requires excited-state methods beyond DFT. The state-of-the-art methodology for extended systems is provided by many-body perturbation theory (MBPT), taking DFT results as a starting point. One- and two-particle Green functions are employed to obtain the quasi-particle eigenvalues and optical spectra accounting for electron-hole interaction in the excitation process, respectively. Here we focus on the Bethe-Salpeter equation of MBPT -- which is typically used for computing absorption spectra in the visible and UV range of light -- to explore core excitations. We sketch the theoretical background and its implementation in a full potential all-electron framework, as realized in the \exciting\ code. For a broad range of materials, we highlight the physics behind the particular spectral features, or the added value compared to experimental investigations. In all cases, we refer to the original literature for further reading.

\section*{\exciting\ -- AN ALL-ELECTRON FULL-POTENTIAL CODE}

\exciting\ is an electronic-structure package that solves the Kohn-Sham (KS) equation of DFT using linearized augmented planewaves (LAPW) as basis set.  By {\it LAPW} we mean various ways of linearizing the KS eigenvalue problem, including their extension by local orbitals of any kind~\cite{gula+14jpcm}. Beyond ground-state DFT methods, \exciting\ implements time-dependent DFT as well as MBPT for treating excitations, the latter comprising the $GW$ approach to obtain quasi-particle bands and the Bethe-Salpeter equation that allows for including excitonic effects in the description of spectra. Besides its capability to compute optical absorption, Raman scattering, and electron loss, \exciting\ has a particular focus on core-level excitations. Owing to the fact that the LAPW method does not require any shape approximation for the potential, density or wavefunction, it can describe all electrons on the same footing, and excitations from any edge can be handled explicitly. More specifically, at present, core-electrons are treated by the solution of the Dirac equation, while the standard approach for valence states is a scalar-relativistic solution of the KS equation, with the option to include spin-orbit interaction in a second-variational scheme. A fully relativistic treatment with spinor wavefunctions is currently under development. 

\section{Methodology}


The Bethe-Salpeter equation is an effective two-body equation for the electron-hole interaction, based on a two-particle Green functions approach. It can be written in terms of an eigenvalue problem in matrix form:
\begin{equation}
\sum\limits_{c'u'\mathbf{k'}} \hat{H}^{e-h}_{cu\mathbf{k},c'u'\mathbf{k'}} A^{\lambda}_{c'u'\mathbf{k'}} = 
E^{\lambda} A^{\lambda}_{cu\mathbf{k}}
\label{eq:BSE}
\end{equation}
In the context of x-ray spectroscopy, we consider only transitions from core ($c$) to unoccupied ($u$) states; $\mathbf{k}$ and $\mathbf{k'}$ are wavevectors of the first Brillouin zone. 

The two-body Hamiltonian consists of three terms, $\hat{H}^{e-h} = \hat{H}^{diag} + \hat{H}^x + \hat{H}^{dir}$. The \textit{diagonal} term $\hat{H}^{diag}$ describes single-particle transitions. Considering only this term corresponds to the independent-particle approximation (IPA). The \textit{exchange} term, $\hat{H}^x$, accounts for local-field effects, while the \textit{direct} term, $\hat{H}^{dir}$, incorporates the attractive screened Coulomb interaction $W$. In the adopted planewave representation, the latter is expressed by $W_{\mathbf{G}\mathbf{G}'}(\mathbf{q})=\frac{4\pi \epsilon^{-1}_{\mathbf{G}\mathbf{G}'}}{|\mathbf{q}+\mathbf{G}||\mathbf{q}+\mathbf{G}'|}$. More details on the BSE formalism and its implementation in the LAPW framework can be found elsewhere~\cite{pusc-ambr02prb,Sagmeister2009,vorw+17prb}. The eigenvalues, $E^{\lambda}$, represent exciton energies, while the eigenvectors, $A^{\lambda}$, carry the information about their character and composition. They enter the imaginary part of the macroscopic dielectric function, $\mathrm{Im}\varepsilon_M$, renormalizing the oscillator strength as compared to the IPA. With the the momentum operator {\bf p} accounting for the dipole selection rules, the diagonal components read:
\begin{equation}
\mathrm{Im}\varepsilon_M = \frac{8\pi^2}{\Omega} \sum\limits_{\lambda cu\mathbf{k}}
\left | A^{\lambda}_{cu\mathbf{k}}\frac{{\langle c\mathbf{k}  |
{\mathbf{p}}|u\mathbf{k}\rangle}}{{\epsilon_{u\mathbf{k}} - \epsilon_{c\mathbf{k}}}} \right|^2 \delta(\omega - E_{\lambda}) .
\label{eq:ImeM}
\end{equation}
This is the quantity used for comparison with experimental spectra and shown below for prototypical materials. 

All the above expressions are evaluated by using the LAPW basis set. This dual basis is based on the partitioning of the unit cell into {\it atomic spheres} around the nuclei and an {\it interstitial region} in between. For details we refer the reader to~\cite{pusc-ambr02prb,Sagmeister2009,gula+14jpcm,ambr-sofo06cpc}.

Before showing the performance of this method in terms of various examples, we would like to add the following notes: 
\begin{itemize}
\item 
{The BSE improves over the {\it super-cell core-hole} approach in particular for excitations from semicore states where electron-hole correlation becomes significant~\cite{olov+09jpcm,rehr+05ps}, but also in the sense that the electron-hole binding does not depend on the approximate functional employed for the ground-state calculation [see example of LiF and Ref.~\cite{olov+09prb}].}
\item 
{So far, the core excitations based on the BSE have often relied on the description of the core states by the help of local orbitals (added to the valence range) rather than accounting for the spinor wavefunctions obtained from the solution of the Dirac eqution; exceptions being~\cite{lask-blah10prb,vorw+17prb}. The latter becomes critical in case of small $2p_{1/2}$ / $2p_{3/2}$ splitting (see example of TiO$_2$ below).}
\item 
{In contrast to optical spectra, the $GW$ step for obtaining the quasi-particle energies is typically avoided when it comes to core excitations, since the $GW$ results do not match core levels well. Therefore the spectra are typically aligned with respect to experiment. The quasi-particle correction applied to the conduction states [see, for example,~\cite{vins+11prb,vins+12prb}] may, however, improve the spectra, especially when the unoccupied bands exhibit different self-energies.}
\end{itemize}

\section{Selected systems}

\subsection{Graphene - an unusual metal}
\begin{figure}
\includegraphics[width=.45\textwidth]{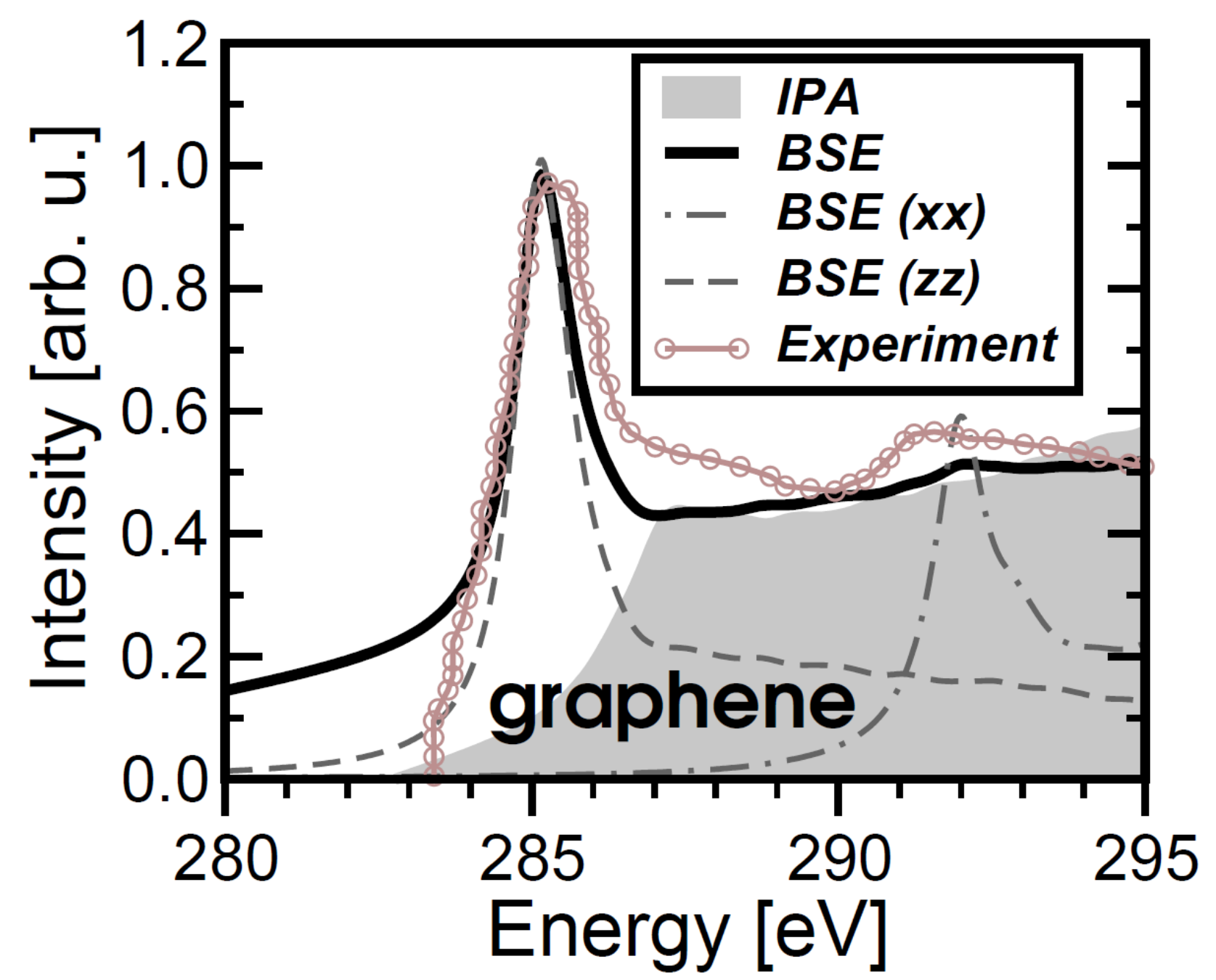}
\caption{Near-edge x-ray absorption spectrum from the carbon $K$-edge of graphene,including the absorbance at grazing incidence of 16$^{\circ}$ (solid line) as well as the in-plane ($xx$, dashed-dotted line) and out-of-plane ($zz$, dashed line) components of the dielectric tensor} obtained by the BSE. The absorbance at grazing incidence from the IPA (shaded area) as well as the experimental data (circles) are shown for comparison.
\label{fig:gr}
\end{figure}

The x-ray absorption spectrum from the carbon $K$-edge (Fig. \ref{fig:gr}) is characterized by a sharp near-edge peak, corresponding to the resonance between the C 1$s$ electron and the lowest unoccupied $\pi^*$ orbital, followed by intense features at higher energies. The measured spectrum~\cite{paci+08prl} is very well reproduced by the BSE approach, while the IPA is in qualitative disagreement~\cite{NOMAD-graphene}. This result indicates that even in a metallic material, like graphene, the Coulomb interaction between the core-hole and the conduction electrons can be relevant. It should be noted that a quantitative description of the experimental features is only obtained by taking into account the laboratory setup. In the experiment~\cite{paci+08prl}, the impinging beam is at grazing incidence (16$^{\circ}$) with respect to the sample. This configuration is reproduced by calculating the absorbance with {\LO}~\cite{vorw+16cpc}, a computational tool for solving Fresnel's equations for anisotropic layered materials~\cite{yeh80ss,pusc-ambr06aem}. For comparison, we also plot in Fig. \ref{fig:gr} the in-plane ($xx$) and out-of-plane ($zz$) components of the macroscopic dielectric function obtained from BSE. It is evident that the shoulder appearing in the experimental spectrum at about 292 eV comes from the peak in the $xx$ component of Im$\epsilon_M$, which enters the absorbance at grazing incidence.

\subsection{LiF -- a wide-gap material}

\begin{figure}
\includegraphics[width=.45\textwidth]{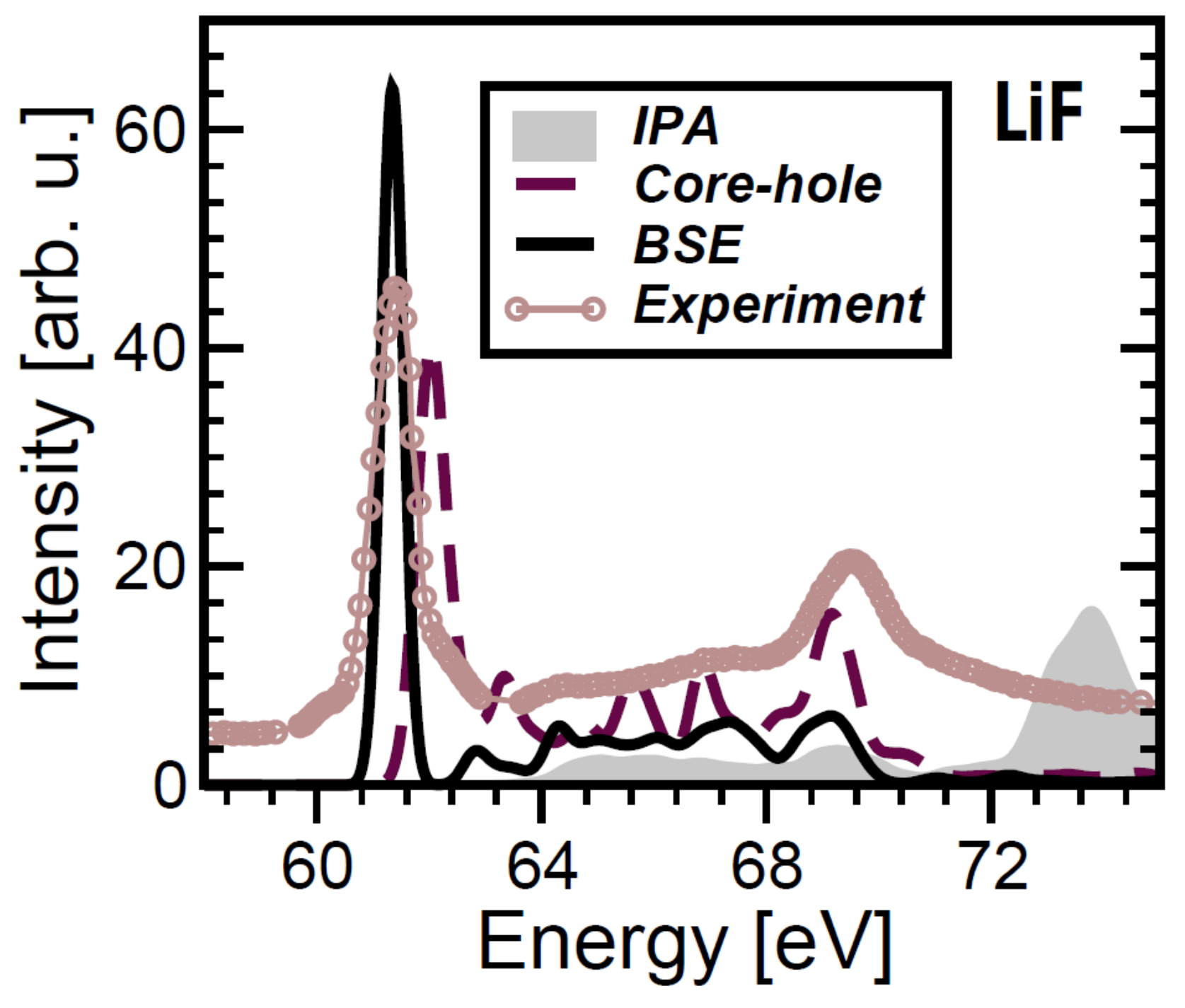}
\caption{Near-edge x-ray absorption spectrum from the lithium $K$-edge in LiF, computed by the BSE (solid line), the core-hole approximation (dashed line), and the IPA (shaded area), in comparison with experimental results (circles). Data from~\cite{olov+09jpcm}.}
\label{fig:LiF}
\end{figure}

LiF exhibits dominant excitonic features, as demonstrated by the Li $K$-edge absorption spectrum shown in Fig. \ref{fig:LiF}. Although the Li 1$s$ electron is a very shallow core state, the importance of core-hole effects is already evident by the failure of the IPA in reproducing the experimental features~\cite{hand+05}. The sharp peak at the absorption onset is well captured by the BSE as well as by the supercell core-hole approach. However, the latter substantially underestimates the exciton binding energy. As pointed out in Ref.~\cite{olov+09prb}, this can be traced back to the fact that in this method the electron-hole interaction is handled on the level of the exchange-correlation functional of the underlying DFT calculation, in contrast to the BSE, where the Coulomb interaction is treated explicitly.

\subsection{AlN -- fingerprints of structural phases}
\begin{figure}
\includegraphics[width=.45\textwidth]{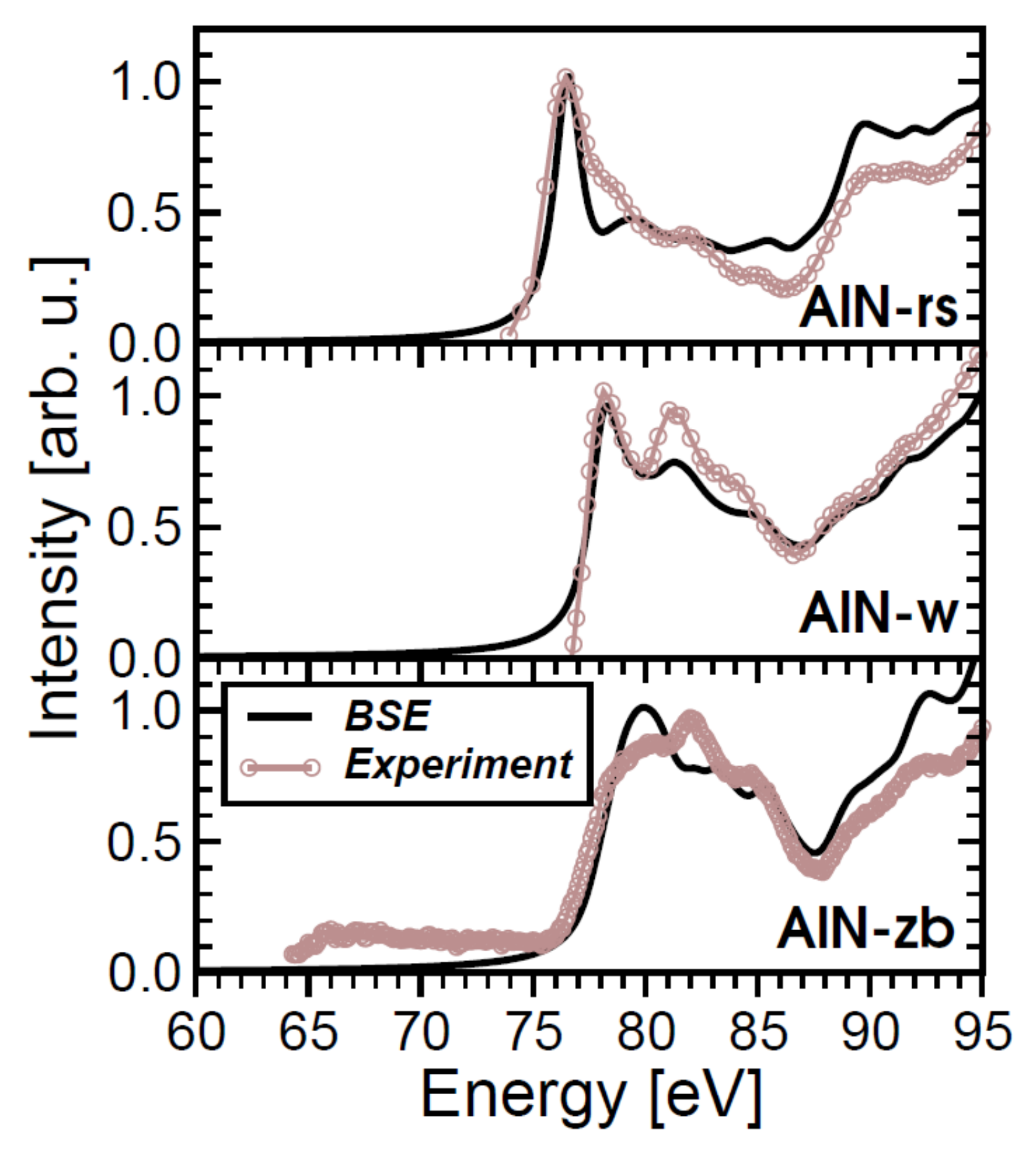}
\caption{Near-edge x-ray absorption spectra from the aluminum $L_{2,3}$ edge of AlN in the rock-salt (rs), wurzite (w), and zinc-blende (zb) phases. Results obtained from the BSE (solid line) are compared with experiment (circles). Data from~\cite{olov+11prb}.}
\label{fig:AlN}
\end{figure}
The NEXAFS spectra of rock-salt (rs), wurzite (w), and zinc-blende (zb) AlN from the aluminum $L_{2,3}$ edge are shown in Fig. \ref{fig:AlN}, comparing theory with experiments~\cite{mizo+03micron,senn-esno03,olov+11prb}. The calculated results impressively demonstrate the power of the BSE to correctly predict the particular features owing to different atomic configurations. While the rock-salt phase shows a sharp peak at the absorption edge, both the wurtzite and the zinkblende structure display a smoother onset. How these features are related to the underlying electronic states and reflected in the nature of the electron-hole wavefunctions of the lowest-energy excitations is discussed in detail in Ref.~\cite{olov+11prb}.

\subsection{Site-selectivity in Ga$_2$O$_3$}

\begin{figure}
\includegraphics[width=.45\textwidth]{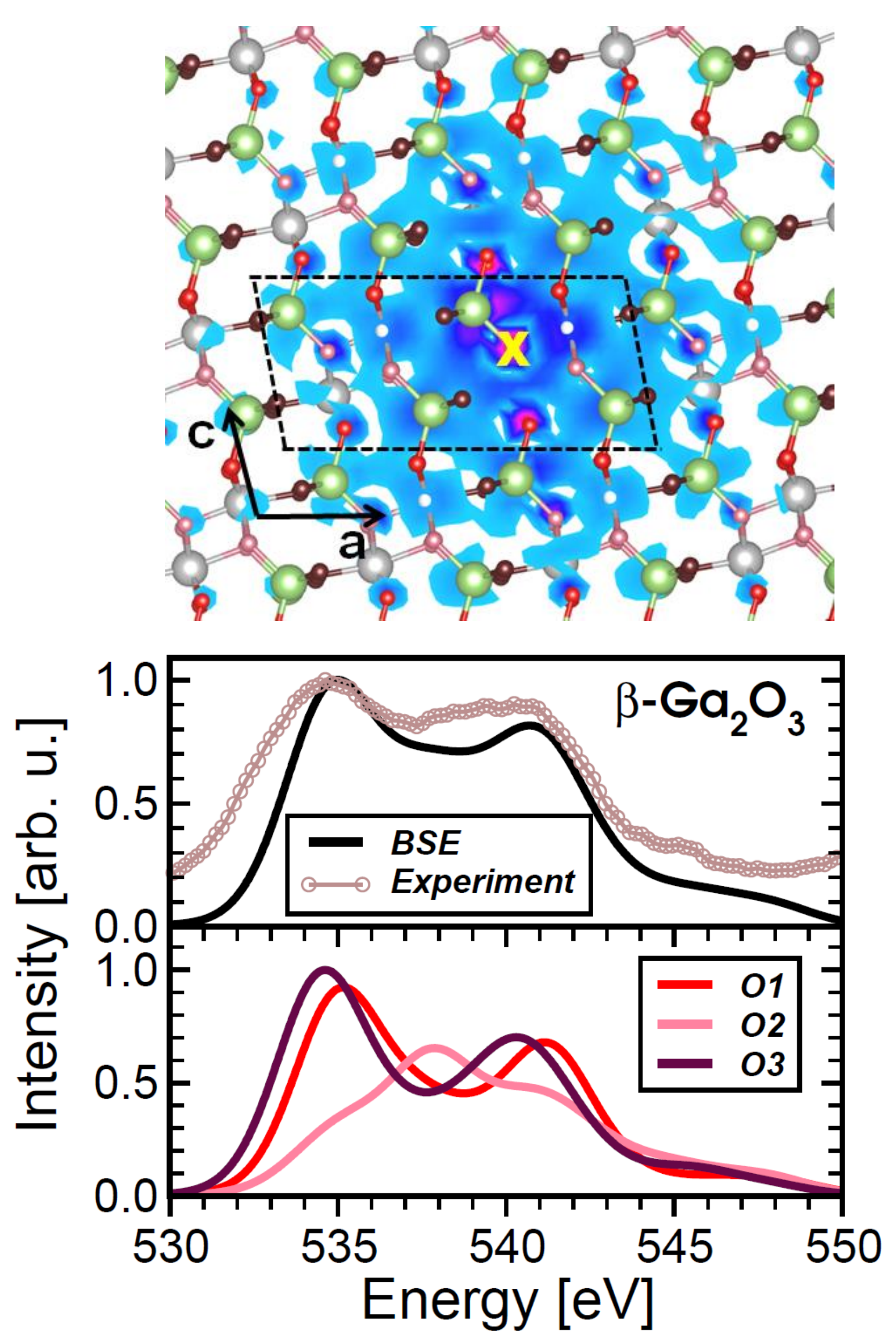}
\caption{Core-excitations from the oxygen $K$-edge in $\beta$-Ga$_2$O$_3$. Top: Electron distribution of the lowest-energy exciton with respect to the hole whose position is marked by the cross. The dashed line indicates the unit cell. Oxygen atoms in red, gallium atoms in grey and green. Bottom (upper panel): BSE results (solid line) and experimental ELNES spectra (circles) obtained for a sample impinged by the electron beam along the direction normal to the (200) surface; lower panel: BSE spectra for inequivalent oxygen atoms. Results from~\cite{cocc+16prb}.}
\label{fig:Ga2O3}
\end{figure}

The x-ray absorption spectrum from the O $K$-edge of $\beta$-Ga$_2$O$_3$ (Fig. \ref{fig:Ga2O3}, bottom) is dominated by excitonic effects which are responsible for the overall spectral shape~\cite{cocc+16prb,NOMAD-ga2o3}. Atom-resolved contributions from chemically inequivalent oxygen atoms of the monoclinic unit cell reveal signatures of their chemical environment that can be probed by ELNES under different diffraction conditions. The corresponding experimental setup is mimicked by theory by rotating the bulk dielectric tensor such to represent the orientation of the sample with respect to the incident electron beam. This is accomplished by {\LO}~\cite{vorw+16cpc}.

These results suggest that ELNES, together with \textit{ab initio} many-body theory, can be successfully employed to characterize complex systems, with sensitivity to individual atomic species and their local environment.

It is also instructive to inspect the real-space extension of the exciton wavefunction (Fig. \ref{fig:Ga2O3}, top).
The shape reflects the electron density related to the bottom of the conduction band, that is formed by hybridized O $p$ and Ga $s$ states.

\subsection{Packing effects in azo-benzene SAMs}

\begin{figure}
\includegraphics[width=.4\textwidth]{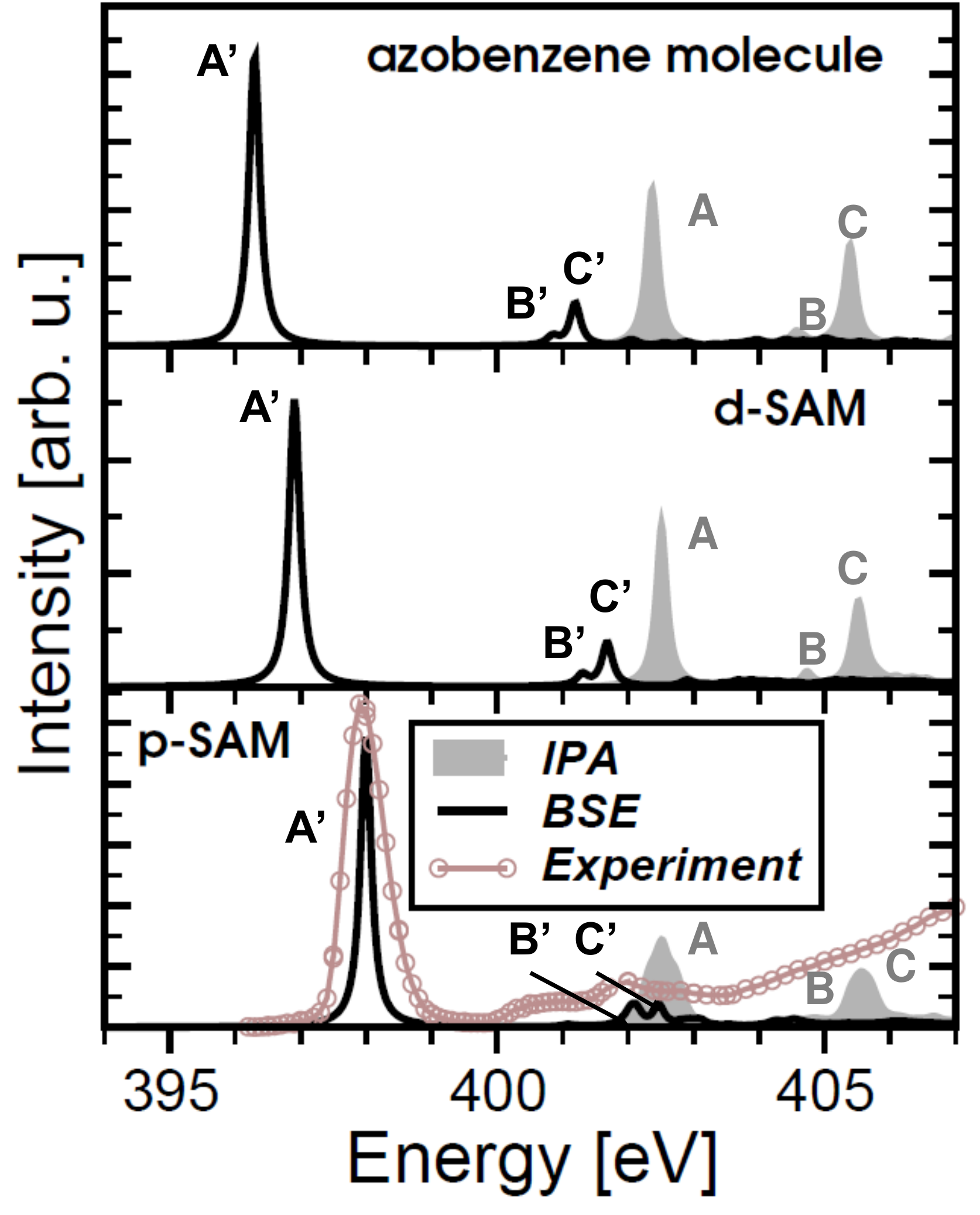}
\caption{Near-edge x-ray absorption spectra from the nitrogen $K$-edge of an isolated azobenzene molecule (top), a diluted SAM (d-SAM, middle), and a packed SAM (p-SAM, bottom). The excitations labeled as A', B' and C' in the BSE spectra (solid lines), can be traced back to peaks A, B, and C, respectively, in the IPA spectra (shaded areas). Theoretical results taken from~\cite{cocc-drax15prb}, experimental data from~\cite{mold+15lang}.}
\label{fig:AB}
\end{figure}

Core excitations from the N $K$-edge in azobenzene molecules and functionalized self-assembled monolayers (SAMs) at 
increasing packing density reveal intermolecular coupling in the excited state~\cite{cocc-drax15prb,NOMAD-az}.
Comparison between BSE and IPA results (Fig. \ref{fig:AB}) highlight the huge exciton binding energies characteristic of organic molecular systems. They range from up to 6 eV in isolated molecules to 4 eV in the densely-packed SAM, where the screening is enhanced by molecular packing. Our results, being in very good agreement with experiment~\cite{mold+15lang}, indicate the crucial role of the electron-hole attraction in determining these excitations.

\subsection{Spin-orbit coupling in TiO$_2$}

\begin{figure}
\includegraphics[width=.42\textwidth]{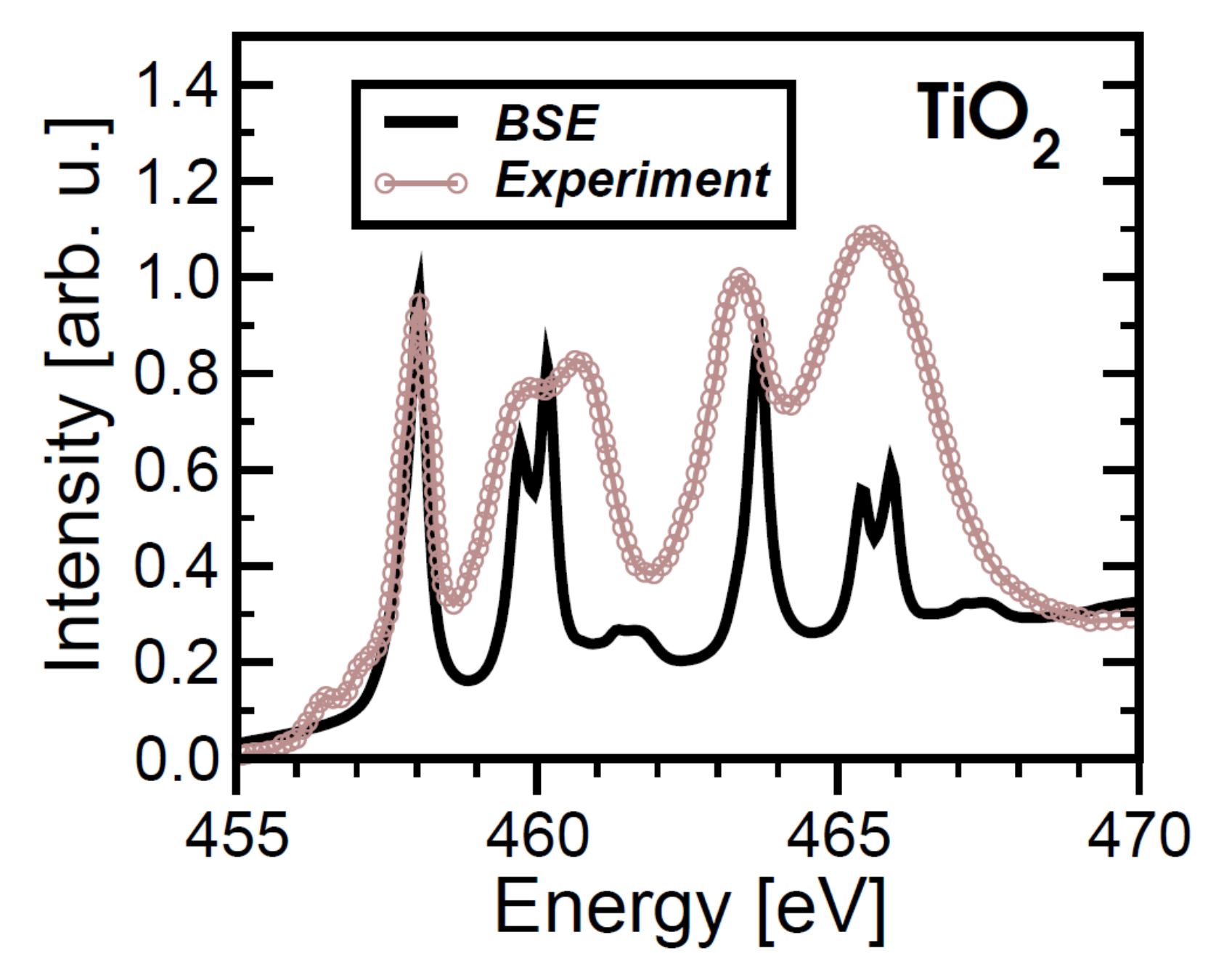}
\caption{Near-edge x-ray absorption spectrum from the titanium $L_{2,3}$-edge in rutile TiO$_2$, computed from the BSE (solid line) and compared with experimental data (circles) taken from~\cite{kron+12prb}.}
\label{fig:TiO2}
\end{figure}

Rutile TiO$_2$ is a prototypical system for studying core-level excitations from the $L_{2,3}$ edge of the metal ion. 
To capture multiplet effects, which are responsible for mixing the single-particle transitions from the $2p_{1/2}$ and $2p_{3/2}$ core states, a fully relativistic treatment together with the explicit electron-hole interaction is essential. Due to the rather small spin-orbit splitting on the order of a few eV, the spectra arising from the $L_2$ and $L_3$ edge cannot be treated separately, as also indicated in Ref.~\cite{lask-blah10prb}. We find the screened Coulomb attraction as well as the repulsive exchange interaction between the excited electron and the core hole to be crucial in order to reproduce the correct spectral features shown in Fig. \ref{fig:TiO2}~\cite{vorwerk-master}.

\section{Summary and outlook}

The BSE is a powerful tool to explore core excitations from first principles, as demonstrated by examples that represent materials and excitations of very different character. Many-body theory is essential to complement experiments, gaining information that is inaccessible by measurements only. For a more stringent interpretation of experiments, further steps are required. These concern the incorporation of electron-phonon coupling as well as higher-order processes beyond the BSE. Also precise knowledge on the sample quality and experimental conditions are a must for quantitative comparison. Improvements and extensions of the \exciting\ code concerning momentum-transfer~\cite{fuga+15prb} as well as spin treatment are already underway. The latter is based on a fully relativistic description of core and valence states, covering also magnetic materials and dichroism. 

\section*{Acknowledgement}
We appreciate funding from the DFG (CRC 658 and CRC 951) and the Leibniz ScienceCampus GraFOx. C.~C.~ acknowledges support from IRIS Adlershof.


\end{document}